# Even-Odd Layer-Dependent Anomalous Hall Effect in Topological Magnet MnBi$_2$Te$_4$ Thin Films

Yi-Fan Zhao[1], Ling-Jie Zhou[1], Fei Wang[1], Guang Wang[1], Tiancheng Song[2], Dmitry Ovchinnikov[2], Hemian Yi[1], Ruobing Mei[1], Ke Wang[3], Moses H. W. Chan[1], Chao-Xing Liu[1], Xiaodong Xu[2,4], and Cui-Zu Chang[1]

[1]Department of Physics, The Pennsylvania State University, University Park, PA 16802, USA

[2]Department of Physics, University of Washington, Seattle, WA 98195, USA

[3]Material Research Institute, The Pennsylvania State University, University Park, PA 16802, USA

[4]Department of Material Science and Engineering, University of Washington, Seattle, WA 98195, USA

Corresponding authors: cxc955@psu.edu (C.-Z. C.).

**Abstract:** A central theme in condensed matter physics is to create and understand the exotic states of matter by incorporating magnetism into topological materials. One prime example is the quantum anomalous Hall (QAH) state. Recently, MnBi$_2$Te$_4$ has been demonstrated to be an intrinsic magnetic topological insulator and the QAH effect was observed in exfoliated MnBi$_2$Te$_4$ flakes. Here, we used molecular beam epitaxy (MBE) to grow MnBi$_2$Te$_4$ films with thickness down to 1 septuple layer (SL) and performed thickness-dependent transport measurements. We observed a non-square hysteresis loop in the antiferromagnetic state for films with thickness greater than 2 SL. The hysteresis loop can be separated into two AH components. Through careful analysis, we demonstrated that one AH component with the larger coercive field is from the dominant MnBi$_2$Te$_4$ phase, while the other AH component **with the smaller coercive field is from the minor Mn-doped $Bi_2Te_3$ phase in the samples. The extracted AH component of the $MnBi_2Te_4$ phase shows a clear even-odd layer-dependent behavior, a signature of antiferromagnetic thin films. Our studies reveal insights on how to optimize the MBE growth conditions to improve the quality of $MnBi_2Te_4$ films, in which the QAH and other exotic states are predicted.**

**Main text:** The quantum anomalous Hall (QAH) effect, a zero magnetic field manifestation of the integer quantum Hall (QH) effect, has attracted a great deal of attention in the condensed matter and material science communities in the past ten years[1-3]. The dissipationless chiral current that flows along the edges of the QAH sample, known as chiral edge current, opens the door to electronics, spintronics, and topological quantum computations with low-energy dissipation[4-6]. However, the QAH effect in magnetically doped topological insulator (TI) systems relies on magnetic dopants and hence introduces disorders into the TI films which inevitably degrade sample quality. The QAH effect in magnetically doped TI systems is found only below ~2 K[3, 7-9]. This low working temperature is the main obstacle for exploring new physics and accessing potential applications of the QAH phenomenon. Alternative QAH platforms are required to overcome the above fundamental material limitations to realize practical quantum technologies. An ideal platform for the realization of the QAH state would be a single crystalline material that hosts coexisting topological and magnetic states intrinsically.

$MnBi_2Te_4$, a tetradymite-type compound, was recently predicted[10, 11] and experimentally demonstrated[12, 13] to be a TI with intrinsic antiferromagnetism. In $MnBi_2Te_4$, a layered antiferromagnetic (AFM) order due to the Mn sublattice emerges below its *Néel* temperature $T_N$ of ~24 K[13-15]. The Mn moments couple ferromagnetically within each septuple layer (SL), but antiferromagnetically between adjacent SLs (Fig. 1a). This A-type AFM order in $MnBi_2Te_4$ has a substantial influence on its topological properties[10, 11, 16]. Because of the ferromagnetism within one SL, each helical surface state is expected to experience a magnetic exchange gap due to surface magnetization, thus contributing half-quantized Hall conductance from the magnetic exchange gap. Depending on whether the MnBi$_2$Te$_4$ film has an even or odd number of SLs, the top and bottom surface magnetizations will be anti-parallel or parallel and the Hall conductance from the two surfaces will cancel or sum. The total conductance is $e^2/h$ in odd SLs (i.e., the QAH state) and 0 in even SLs (i.e., the "axion insulator" state) under zero magnetic field [10, 11, 16]. The QAH state under zero magnetic field (i.e. the AFM state) and the Chern insulator state under high magnetic fields (i.e. the ferromagnetic state) have recently been realized in mechanically exfoliated MnBi$_2$Te$_4$ flakes[17-21]. However, the exfoliated flakes are fragile and typically small with irregular shapes and thus prohibit the large-scale fabrication of technological devices that make use of the QAH effect. Therefore, the molecular beam epitaxy (MBE) growth of high-quality MnBi$_2$Te$_4$ films is essential. To date, the MBE growth of MnBi$_2$Te$_4$ has been pursued by several groups[12, 22-28], but a systematic thickness-dependent of transport study on MnBi$_2$Te$_4$ films is still lacking.

In this work, we performed systematic electrical transport measurements on MBE-grown films with thickness down to 1 SL and observed a non-square hysteresis loop in the AFM regime of the MnBi$_2$Te$_4$ films with thickness greater than 2 SL. This non-square hysteresis loop can be separated into two AH components and shown to originate from two coexisting phases in the MBE-grown MnBi$_2$Te$_4$ films. One is from the dominant MnBi$_2$Te$_4$ phase with a larger coercive field, the other is from the minor Mn-doped Bi$_2$Te$_3$ phase with a smaller coercive field. We also found that the extracted AH effect of the MnBi$_2$Te$_4$ phase shows a clear even-odd layer-dependent behavior. The coexistence of MnBi$_2$Te$_4$ and Mn-doped Bi$_2$Te$_3$ phases might account for the small Hall resistance in MBE-grown MnBi$_2$Te$_4$ films[12, 22, 27, 28].

All MnBi$_2$Te$_4$ films were grown on ~0.5 mm thick heat-treated SrTiO$_3$(111) substrates in a MBE chamber (Omicron Lab 10) with a base vacuum ~2 × 10$^{-10}$ mbar. The *in-situ* angle-resolved photoemission spectroscopy (ARPES) measurements were carried out using a Scientia R3000 analyzer with an unpolarized He-Iα light (~21.2 eV). The electrical transport studies were carried out in a Physical Property Measurements System (Quantum Design DynaCool, 2 K, 9 T) with the magnetic field applied perpendicular to the film plane. Six terminal mechanically defined Hall bars were used for electrical transport measurements. More details about the sample synthesis, ARPES, and transport measurements can be found in Experimental Methods.

We first characterized the MnBi$_2$Te$_4$ films grown on SrTiO$_3$(111) substrates. The sharp and streaky "1×1" reflection high-energy electron diffraction (RHEED) patterns (Fig. S1) and sharp X-ray diffraction (XRD) peaks (Fig. S2) demonstrate the highly ordered and smooth surface of the MnBi$_2$Te$_4$ films. The high-resolution scanning transmission electron microscopy (STEM) measurements show that MnBi$_2$Te$_4$ film is comprised of a SL of Te-Bi-Te-Mn-Te-Bi-Te (Fig. 1b). The thickness of each SL from the XRD data is ~1.36 nm, consistent with literature [12, 14, 17, 18, 20]. Figure 1c shows the *in-situ* ARPES band map of 20 SL MnBi$_2$Te$_4$ film. A ~150 meV gap was observed up to room temperature. Since the observed gap can still be detected even above $T_N$, it might not be related to the AFM order in MnBi$_2$Te$_4$. The observed gap is consistent with the band structures in prior ARPES studies on MnBi$_2$Te$_4$ single crystals [13, 14, 29] but different from the gapless surface states in reported MBE-grown MnBi$_2$Te$_4$ films measured with the same photon energy (i.e. $hv$ ~ 21.2 eV)[12]. For MnBi$_2$Te$_4$ single crystals, gapless surface states are observed with photon energy $hv \leq 16$ eV but gapped ones for $hv > 16$ eV[30-33]. However, for MBE-grown MnBi$_2$Te$_4$ films, more studies are needed to clarify whether the surface states are gapped or gapless with $hv$ ~ 21.2 eV.

After the ARPES measurements, we capped the 20 SL films with a 10 nm Te layer (see Experimental Methods) and then performed *ex-situ* electrical transport measurements. Figure 1d displays the temperature dependence of the sheet longitudinal resistance $\rho_{xx}$, which shows a metallic behavior above $T \sim 30$ K. A hump feature is observed at $T \sim 24$ K, corresponding to $T_N$.[14, 34] The Hall trace of the 20 SL film shows a clear deviation from the linear ordinary Hall curve (i.e. a kink feature) at magnetic field $\mu_0 H \sim 3.7$ T, which corresponds to the occurrence of the spin-flop transition between AFM and canted AFM state in the metallic regime[34].

In the following, we will focus on transport measurements on MnBi$_2$Te$_4$ films with thickness from 1 SL to 7 SL. Figure 2 shows the temperature dependence of $\rho_{xx}$ under zero magnetic field. For film thickness less than 5 SL, $\rho_{xx}$-$T$ shows an insulating behavior. However, for the 6 SL and 7 SL samples, $\rho_{xx}$-$T$ shows a metallic behavior. For the samples with thickness greater than 3 SL, $\rho_{xx}$-$T$ also show hump features near the *Néel* temperature $T_N$, which allow us to determine $T_N$s of $\sim 20$ K, $\sim 21$ K, $\sim 22$ K, $\sim 23$ K, and $\sim 23.7$ K for the 3, 4, 5, 6, and 7 SL samples, respectively. The higher magnetic ordering temperature in thicker films is similar to that observed in the magnetically doped TI films[35]. The observation of the $T_N$ hump feature with film thickness down to 3 SL demonstrates the dominance of the MnBi$_2$Te$_4$ phase, implying the high quality of our thin films. We note that the $T_N$ hump feature is absent in all MnBi$_2$Te$_4$ films with thickness less than 10 SL in prior reports[12, 22, 27, 28].

After characterizing the temperature dependence of $\rho_{xx}$ under zero magnetic field, we carried out the corresponding Hall measurements. For the 1 SL film, $\rho_{xx} \sim 350$ kΩ at $T \sim 2$ K, which makes the Hall measurements unreliable. Therefore, we focus on the Hall measurements with film thickness greater than 2 SL. Figures 3a to 3f show the magnetic field $\mu_0 H$ dependence of the Hall resistance $\rho_{yx}$ from 2 SL to 7 SL at $T$ =2 K. The small negative slope of the Hall traces indicate all these MnBi$_2$Te$_4$ films are heavily electron-doped, i.e. the Fermi level is located within the bulk conduction bands, consistent with our ARPES band maps (Fig. S5). We found that the Hall traces of the MBE-grown metallic MnBi$_2$Te$_4$ films show the typical kink feature (marked by the arrows in Figs. 3a to 3f) in the $\mu_0H$ range from ~2.5 T to 4.0 T[34]. As noted above, this kink feature corresponds to the occurrence of the spin-flop transition in the samples. Moreover, our observation also suggests that the A-type AFM order persists down to the 2 SL MnBi$_2$Te$_4$ film. We deduce the spin-flop magnetic field $\mu_0H_1$ from the kink feature in $\rho_{yx}$ and a peak feature in $\rho_{xx}$ (Fig. S7), which are ~2.3 T, ~3.75 T, ~3.5 T, ~4 T, ~3.65 T, and ~3.8 T for the 2, 3, 4, 5, 6, and 7 SL samples, respectively. These values are slightly different from $\mu_0H_1$s from our reflection magnetic circular dichroism (RMCD) measurements on the same samples (Fig. S6). We note that the lower $\mu_0H_1$ in even SL but higher $\mu_0H_1$ in odd SL heavily electron-doped MnBi$_2$Te$_4$ films grown by MBE are consistent with two recent RMCD reports on mechanically exfoliated MnBi$_2$Te$_4$ flakes[20, 36]. Compared to thicker MBE-grown MnBi$_2$Te$_4$ films, the much lower $\mu_0H_1$ in the 2 SL sample might be caused by its weaker anisotropy.

In the low magnetic field regime corresponding to the AFM state, a non-square hysteresis loop was observed in all samples, which is especially prominent for the 2 SL sample and gradually becomes less appreciable as film thickness increases (see the insets of Figs. 3a to 3f for details). Note that for the 6 SL and 7 SL samples, the Hall trace is similar to that of the 20 SL film (Fig. 1d inset), but a tiny non-square hysteresis is still present in zoomed-in curves near zero magnetic field. The non-square AH hysteresis loop observed in the AFM state shows a striking even-odd layer number dependent behavior. In even SL samples, hysteresis loops have a hump feature, while in odd SL samples, it becomes a two-step magnetic transition feature. This observation may be induced by the superposition of two AH effects with opposite[37, 38] and same signs [39, 40] for the even and odd layer numbers, respectively, which will be examined below.

We expressed $\rho_{yx}$ of the MnBi$_2$Te$_4$ films as: $\rho_{yx} = \rho_{OH} + \rho_{AH1} + \rho_{AH2}$. Here $\rho_{OH}$ is the ordinary Hall effect, $\rho_{AH1}$ and $\rho_{AH2}$ are the two AH components that reside in the observed non-square AH hysteresis loop. To extract $\rho_{AH1}$ and $\rho_{AH2}$, we subtracted $\rho_{OH}$ from $\rho_{yx}$ to resolve the AH resistance $\rho'_{yx}$, which is then fitted with the below equation[41]:

$$\rho'_{yx} = \rho_1 \tanh(w_1(H - H_{C1})) + \rho_2 \tanh(w_2(H - H_{C2}))$$

Here $\rho_1$($\rho_2$), $H_{C1}$($H_{C2}$) are the amplitude and the coercive field of the first (second) AH component $\rho_{AH1}$ ($\rho_{AH2}$), respectively; $w_1$ and $w_2$ are two constants. The fitted curves (marked by the dashed lines) are shown in the insets of Figs. 3a to 3f and the extracted $\rho_{AH1}$ and $\rho_{AH2}$ are displayed in Figs. 3g to 3l and Figs. 3m to 3r, respectively.

The extracted $\rho_{AH1}$ shows an even-odd dependent behavior in the AFM regime. For even SL samples, $\rho_{AH1}$ has a positive AH sign (i.e, $\rho_{AH1}>0$ for magnetization $M >0$), and $H_{C1}$s are ~1.5 T, ~1.6 T, and ~1.75 T for the 2 SL, 4 SL, and 6 SL samples, respectively. However, for odd SL samples, $\rho_{AH1}$ has a negative AH sign (i.e, $\rho_{AH1}< 0$ for magnetization $M >0$), and $H_{C1}$s are ~1.0 T, ~1.0 T, and ~1.2 T for the 3 SL, 5 SL, and 7 SL samples, respectively. Therefore, both the AH sign and the coercive field $H_{C1}$ of $\rho_{AH1}$ show an even-odd layer-dependent behavior in the AFM regime of MnBi$_2$Te$_4$ films. Unlike $\rho_{AH1}$, the extracted $\rho_{AH2}$ has the same negative AH sign with a much smaller coercive field $H_{C2}$ ~0.2 T. The $\rho_{AH2}$ under zero magnetic field (labeled as $\rho_{AH2}(0)$) shows a monotonic decrease with increasing film thickness. Taken all these observations together, we speculate that $\rho_{AH1}$ is likely from an AFM phase and $\rho_{AH2}$ from an FM phase in our samples.

To uncover the physical origins of $\rho_{AH1}$ and $\rho_{AH2}$, we performed temperature-dependent Hall measurements. We found that in all samples, the non-square AH hysteresis loop disappears above $T \sim 25$ K (Fig. S13), which is consistent with $T_N$ deduced from the hump features observed in $\rho_{xx}$-$T$ curves (Fig. 2). At each temperature, we separated the non-square hysteresis into two AH components $\rho_{AH1}$ and $\rho_{AH2}$. Figures 4a to 4f show temperature-dependent $\rho'_{yx}$, $\rho_{AH1}$, and $\rho_{AH2}$ of the 3 SL sample. We found that the two-step magnetic transition feature disappears at $T \sim 20$ K, indicating that the magnetic ordering temperature of one of the AH components is lower than $T \sim 20$ K. Since the $H_C$ of the entire hysteresis loop (Fig. 4f) at $T \sim 20$ K is larger than the $H_C$ of the extracted $\rho_{AH2}$ at $T \sim 15$ K, the residual AH hysteresis observed at $T \sim 20$ K should be from $\rho_{AH1}$. The magnetic field dependence of $\rho'_{yx}$, $\rho_{AH1}$, and $\rho_{AH2}$ of other MnBi$_2$Te$_4$ samples at different temperatures are shown in Figs. S8-S12.

To make the even-odd layer-dependent behavior clearer, we plotted the temperature dependence of the zero magnetic field $\rho_{AH1}$ and $\rho_{AH2}$ (labeled as $\rho_{AH1}(0)$ and $\rho_{AH2}(0)$) in Figs. 4g and 4h. $\rho_{AH1}(0)$ of all MnBi$_2$Te$_4$ samples disappear at $T \sim 25$ K, while $\rho_{AH2}(0)$ disappears at $T \sim 20$ K. For the positive magnetization, $\rho_{AH1}(0)$ is positive for the 2 SL, 4 SL, and 6 SL samples but negative for the 3 SL, 5 SL, and 7 SL samples. The magnitude of $\rho_{AH1}(0)$ at $T \sim 2$ K shows a monotonic decrease with increasing film thickness (Fig. 4g). The observation of the even-odd dependent behavior in $\rho_{AH1}(0)$-$T$ curves further confirms that $\rho_{AH1}$ is from an AFM phase, i.e. the dominant MnBi$_2$Te$_4$ phase in our MBE-grown samples.

In all samples, the extracted $\rho_{AH2}(0)$ is always negative and becomes zero at $T \sim 20$K. This further indicates that $\rho_{AH2}$ is from an FM phase with a Curie temperature $T_C$ between 15 K and 20 K. Because the non-square AH hysteresis loop is absent in the metallic regime of the mechanically exfoliated MnBi$_2$Te$_4$ flakes with tens of micrometer size[42], we speculate $\rho_{AH2}$ might be the consequence of inhomogeneity of MBE-synthesized films. Compared to the micrometer size exfoliated MnBi$_2$Te$_4$ devices[17-21, 42], the MnBi$_2$Te$_4$ Hall bar in our transport measurements is ~1 mm × 0.5 mm (see Experimental Methods). For the millimeter size MnBi$_2$Te$_4$ devices, especially for an AFM material, we need to consider the following two aspects below.

The first is that MnBi$_2$Te$_4$ follows an SL-by-SL growth mode[12, 22, 27, 28], which is similar to the quintuple layer (QL)-by-QL growth mode of the well-studied Bi$_2$Se$_3$ family TI[43-45]. Through SL-by-SL growth mode, it is unlikely to achieve the MnBi$_2$Te$_4$ with a uniform thickness in a millimeter size range. Our atomic force microscopy images indeed show that 1 SL islands and holes exist in MBE-grown films (Fig. S3). This property may affect the Hall traces of the AFM MnBi$_2$Te$_4$ films. This can also explain why the extracted $\rho_{AH1}$ of the dominant MnBi$_2$Te$_4$ phase is also slightly different from the Hall traces of the exfoliated MnBi$_2$Te$_4$ devices[17-21, 42]. Moreover, since MnBi$_2$Te$_4$ is crystallized primarily by the annealing process during the MBE growth, the residual Bi$_2$Te$_3$ and MnTe phases and the Mn-doped Bi$_2$Te$_3$ phase may coexist with the dominant MnBi$_2$Te$_4$ phase, which can be seen in some STEM images (Fig. S4). This property becomes more discernible in a millimeter size sample. Based on these two properties, we conclude that $\rho_{AH2}$ is from the Mn-doped Bi$_2$Te$_3$ phase. Further studies are needed to clarify why all MnBi$_2$Te$_4$ films show similar $H_c$ and $T_C$. The small $H_c$ (~ 0.2 T) and low $T_C$ (15~20 K) indicate that the Mn doping concentration in the minor Mn-doped Bi$_2$Te$_3$ phase is at a low level[46, 47].

To summarize, we used MBE to synthesize MnBi$_2$Te$_4$ thin films down to 1 SL and performed thickness-dependent transport measurements. By analyzing the AH hysteresis in the AFM regime of MnBi$_2$Te$_4$ films, we found there is a minor Mn-doped Bi$_2$Te$_3$ phase that coexists with the dominant MnBi$_2$Te$_4$ phase, and the extracted AH traces of the "pure" MnBi$_2$Te$_4$ phase shows an even-odd layer dependent behavior. The existence of a minor FM Mn-doped $Bi_2Te_3$ phase might be responsible for the small Hall resistance usually observed in the MBE-grown $MnBi_2Te_4$ thin films[12, 22, 27, 28]. To realize the QAH/Chern insulator, we need to further optimize the growth recipe to achieve the single $MnBi_2Te_4$ phase (e.g. by fine-tuning of $Bi_2Te_3$/MnTe ratio and the growth/annealing temperatures) and/or fabricate the micrometer size device within a single $MnBi_2Te_4$ phase region. The realization of the QAH/Chern insulator states in wafer-scale MBE-grown $MnBi_2Te_4$ films will advance the fundamental inquiries into topological phases of matter in the presence of both magnetism and band topology and the development of topological materials.

**Experimental Methods**

**MBE growth of $MnBi_2Te_4$ films**

The $MnBi_2Te_4$ films were grown in a commercial MBE system (Omicron Lab 10) with a base vacuum better than $2 \times 10^{-10}$ mbar. All $SrTiO_3$(111) substrates used in our MBE growth were heat-treated in a tube furnace with flowing oxygen[3, 48]. These heat-treated $SrTiO_3$(111) substrates were then loaded into the MBE chamber and outgassed at ~ 500 °C for an hour before the growth of the $MnBi_2Te_4$ films. High purity Mn(99.9998%), Bi(99.9999%), and Te(99.9999%) were evaporated from Knudsen effusion cells. We grew the $MnBi_2Te_4$ films using the method in a prior report [12]. To grow 1 SL $MnBi_2Te_4$ film, we first deposited 1 QL $Bi_2Te_3$ and then deposited 1 BL MnTe on its top. During the growth of these two layers, the substrate is maintained at ~270°C. Next, the heterostructure film was annealed at ~270°C for ~10 minutes to achieve each SL (i.e. Te-Bi-Te-Mn-Te-Bi-Te) structure. Thick $MnBi_2Te_4$ films were achieved by repeating this growth cycle. The growth rates of the $Bi_2Te_3$ and MnTe are ~0.2 QL/min and ~0.6 BL/min, respectively. Finally, the samples were capped with a 10 nm Te layer to prevent their degradation during the *ex-situ* electrical transport measurements and other sample characterizations.

**Electrical transport measurements**

All MnBi$_2$Te$_4$ films for electrical transport measurements were scratched into a Hall bar geometry using a computer-controlled probe station. The effective area of the Hall bar is ~1 mm × 0.5 mm. The electrical contacts were made by pressing indium dots on the films. Transport measurements were conducted using a Physical Property Measurement System (Quantum Design DynaCool, 2 K, 9 T). The excitation current is 1 µA. All magnetotransport results shown in this paper were symmetrized or anti-symmetrized as a function of the magnetic field to eliminate the influence of the electrode misalignment.

***In-situ* ARPES measurements**

The *in-situ* ARPES measurements were performed in a chamber with a base vacuum of ~5×10$^{-11}$ mbar. The MBE-grown MnBi$_2$Te$_4$ films were transferred to the ARPES chamber without breaking the ultrahigh vacuum. The photoelectrons were excited by an unpolarized He-I$_\alpha$ light (~21.2 eV), and a Scientia R3000 analyzer was used for the ARPES measurements. The energy and angle resolutions are ~10 meV and ~0.2°, respectively. All ARPES measurements were performed at room temperature.

**RMCD measurements**

The RMCD measurements on MBE-grown films were performed in a closed-cycle helium cryostat (attoDRY 2100) at *T* ~1.6 K and an out-of-plane magnetic field up to 9 T. A 1000 nm laser was used to probe the samples at normal incidence with the fixed power of ~1 µW. The AC lock-in measurement technique is used to measure the RMCD signals. The experimental setup follows closely with our prior measurements on exfoliated MnBi$_2$Te$_4$ samples[20].

**Supporting Information.** The Supporting Information is available free of charge on the ACS

Publications website.

More sample characterizations, electrical transport results, RMCD data, ARPES band maps, and atomic force microscopy images of MnBi$_2$Te$_4$ films; and the non-square AH hysteresis of more MnBi$_2$Te$_4$ films fitted by the two AH component model.

**Author contributions:** C.-Z. C. conceived and designed the experiment. Y.-F. Z. and L.-J. Z. grew all MnBi$_2$Te$_4$ films and carried out the PPMS transport measurements with the help of C. -Z. C.. Y. -F. Z., H. Y., and G. W. carried out the *in-situ* ARPES measurement with the help of C. -Z. C.. F. W. and K. W. performed the STEM measurements. T. S., D. O. performed the RMCD measurements with the help of X. X.. Y.-F. Z. and C.-Z. C. analyzed the data and wrote the manuscript with inputs from all authors.

**Notes:** The authors declare no competing financial interest.

**Acknowledgments:** We thank Yongtao Cui, Nitin Samarth, Weida Wu, and Di Xiao for the helpful discussion. This work is primarily supported by the AFOSR grant (FA9550-21-1-0177). The ARPES measurements were partially supported by the ARO Young Investigator Program Award (W911NF1810198). The electrical transport measurements and the data analysis are partially supported by the DOE grant (DE-SC0019064). C. Z. C. also acknowledges the support from the Gordon and Betty Moore Foundation's EPiQS Initiative (Grant GBMF9063 to C. -Z. C.).

**Figures and figure captions:**

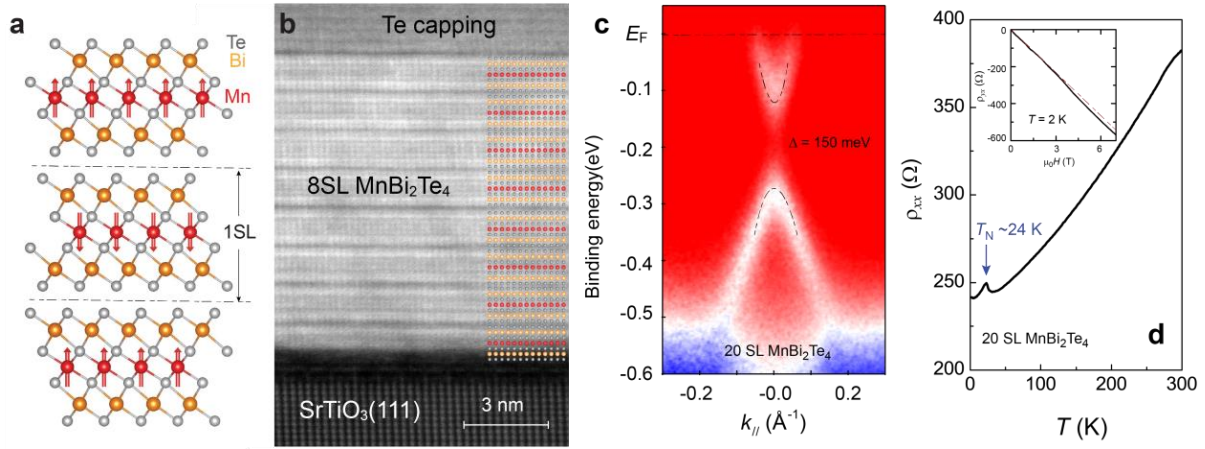

**Figure 1| MBE-grown MnBi$_2$Te$_4$ films on heat-treated SrTiO$_3$(111).** (a) Crystal structure of MnBi$_2$Te$_4$. 1 SL of MnBi$_2$Te$_4$ is composed of Te-Bi-Te-Mn-Te-Bi-Te and the red arrows indicate the magnetic moment of the Mn layer. (b) Cross-sectional STEM image of an 8 SL MnBi$_2$Te$_4$ film grown on SrTiO$_3$(111) substrate. (c) ARPES band map of a 20 SL MnBi$_2$Te$_4$ film grown on SrTiO$_3$(111) substrate. (d) Temperature dependence of the sheet longitudinal resistance ρ$_{xx}$ of the 20 SL MnBi$_2$Te$_4$ film. Inset: the nonlinear Hall trace of the 20 SL MnBi$_2$Te$_4$ film measured at $T$ =2 K. The arrow indicates its *Néel* temperature $T_N$ of ~24 K.

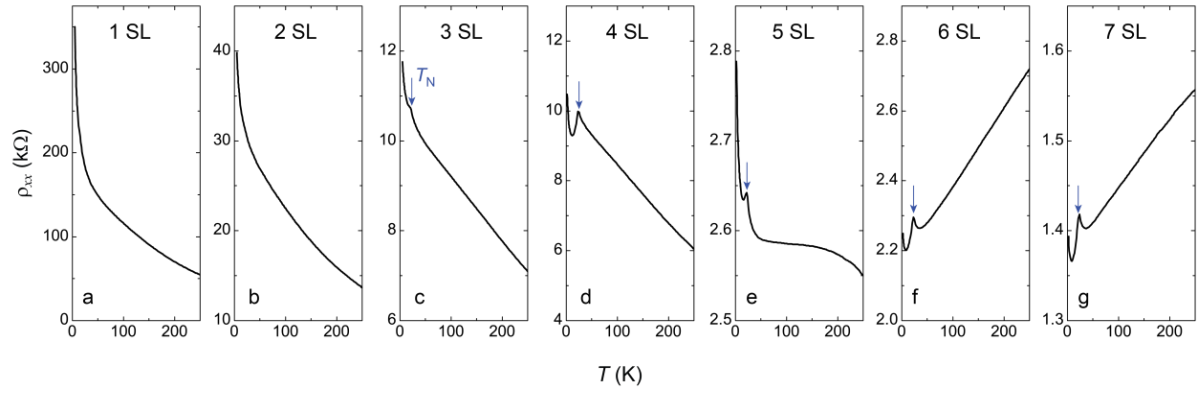

**Figure 2| Thickness dependence of $\rho_{xx}$-$T$ curves of MnBi$_2$Te$_4$ films.** (a-g) Temperature dependent $\rho_{xx}$ of MnBi$_2$Te$_4$ films of different thicknesses. 1 SL(a); 2 SL(b); 3 SL(c); 4 SL(d); 5 SL(e); 6 SL(f); 7 SL(g). The arrows in (c-g) indicate $T_N$s of these MnBi$_2$Te$_4$ films. All measurements were taken at $\mu_0 H$ =0 T.

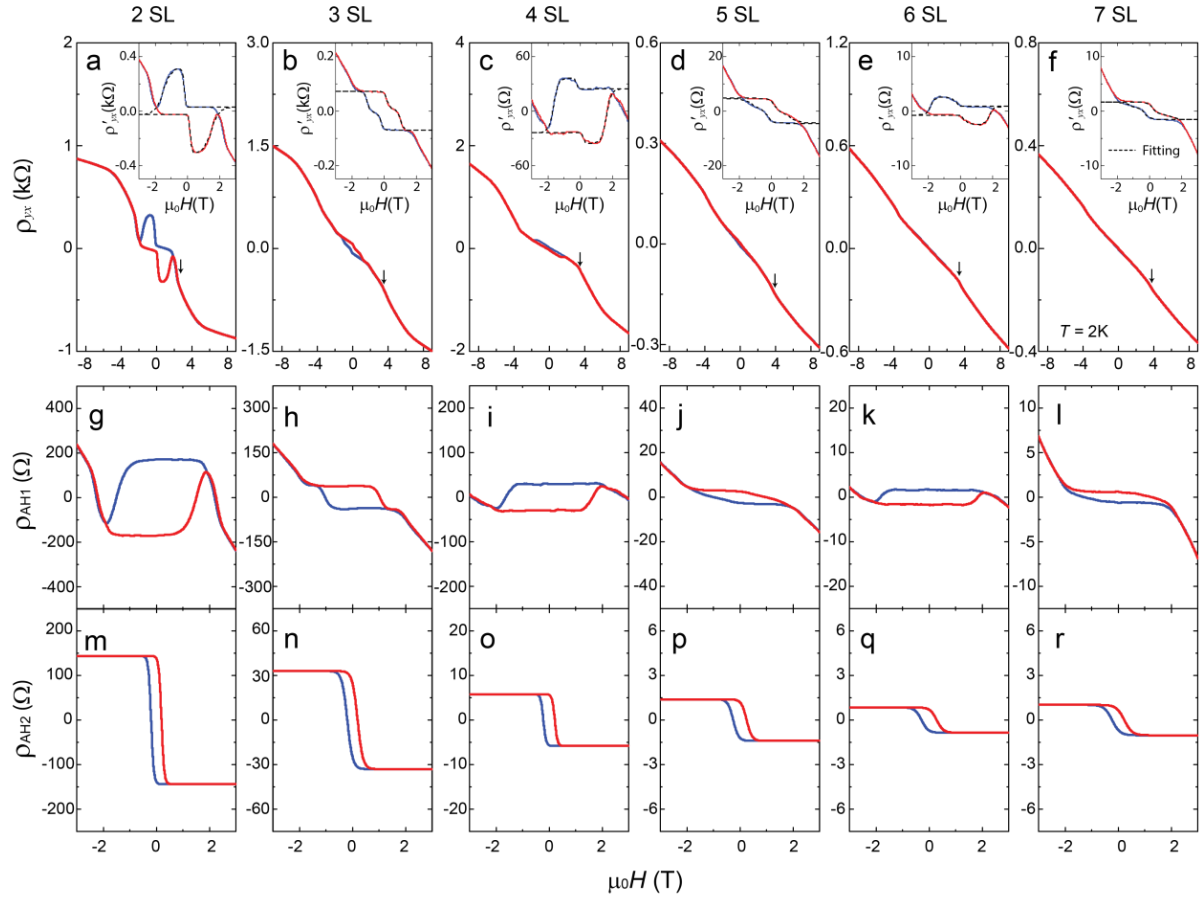

**Figure 3| Thickness dependence of the Hall traces of MnBi$_2$Te$_4$ films measured at *T* =2 K.** (a-f) $\mu_0H$ dependent $\rho_{yx}$ of MnBi$_2$Te$_4$ films of thickness from 2 SL to 7 SL. Inset: The corresponding Hall resistance $\rho'_{yx}$ after subtracting the ordinary Hall effect in the AFM regime. The dashed lines are the curves fitted by the two AH component model, described in the main text. The extracted first (g-l) and second (m-r) AH component $\rho_{AH1}$ and $\rho_{AH2}$ from 2 SL to 7 SL. The blue (red) curve represents the process for decreasing (increasing) $\mu_0H$.

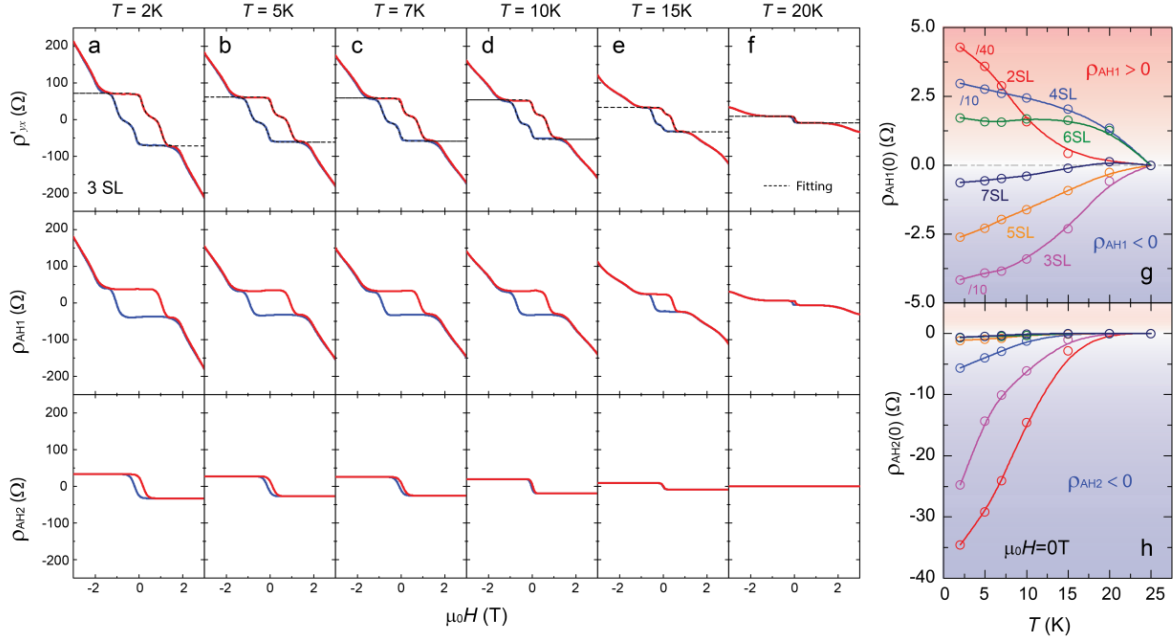

**Figure 4| Magnetic field dependence of the two AH components in MBE-grown MnBi₂Te₄ films at different temperatures.** (a-f) The subtracted AH resistance $\rho'_{yx}$, the first AH component $\rho_{AH1}$, and the second AH component $\rho_{AH2}$ of the 3 SL MnBi₂Te₄ films. $T$ =2 K (a); $T$ =5 K (b); $T$ =7 K (c); $T$ =10 K (d); $T$ =15 K (e); $T$ =20 K (f). The dashed lines in the top panel are the curves fitted by the two AH component model. See the main text for the details. The blue (red) curve represents the process for decreasing (increasing) $\mu_0 H$. See Supporting Information for additional data of other samples. (g, h) Temperature-dependent zero magnetic field $\rho_{AH1}(0)$ and $\rho_{AH2}(0)$ of different layer thicknesses.